\documentclass[conference]{IEEEtran}
\IEEEoverridecommandlockouts
\pdfoutput=1
\usepackage{cite}
\usepackage{amsmath,amssymb,amsfonts}
\usepackage{algorithmic}
\usepackage{graphicx}
\usepackage{textcomp}
\usepackage{enumitem}
\setlist{nosep, leftmargin=14pt}
\usepackage{mwe} 
\usepackage{color}
\usepackage{booktabs}

\usepackage{xcolor}
\def\BibTeX{{\rm B\kern-.05em{\sc i\kern-.025em b}\kern-.08em
    T\kern-.1667em\lower.7ex\hbox{E}\kern-.125emX}}

\begin{document}

\title{Optimized Magnetic Resonance Fingerprinting Using Ziv-Zakai Bound
\thanks{This work was partially supported by the National Key Technology Research and Development Program of China (2023YFC2411103, 2023YFB3811403 and 2023YFF0714201), the National Natural Science Foundation of China(62271474, 62371167), the Natural Science Foundation of Heilongjiang (YQ2021F005), the Fundamental Research Funds for the Central Universities (HIT.OCEF.2024002), the International Partnership Program of Chinese Academy of Sciences (321GJHZ2023246GC), the Guangdong Basic and Applied Basic Research Foundation (2024A1515012138), the High-level Talent Program in Pearl River Talent Plan of Guangdong Province (2019QN01Y986), and the Shenzhen Science and Technology Program (JCYJ20210324115810030 and KJZD20230923113259001).\\
  \hspace*{1em} $^{\ast}$Corresponding authors: Haifeng Wang (email: hf.wang1@siat.ac.cn) and Yue Hu (email: huyue@hit.edu.cn).}
}

\author{\IEEEauthorblockN{Chaoguang Gong$^{\dagger}$, Yue Hu$^{\dagger,\ast}$, Peng Li$^{\dagger}$, Lixian Zou$^{\star }$, Congcong Liu$^{\star }$, \\Yihang Zhou$^{\star }$,Yanjie Zhu$^{\star }$, Dong Liang$^{\star }$, Haifeng Wang$^{\star,\ast}$}
\IEEEauthorblockA{$^{\dagger}$Department of Electronics and Information Technology, Harbin Institute of Technology, Harbin, China\\
 $^{\star }$Paul C. Lauterbur Research Center for Biomedical Imaging, Shenzhen Institutes of \\Advanced Technology, 
 Chinese Academy of Sciences, Shenzhen, Guangdong, China}
}
\maketitle

\begin{abstract}
  Magnetic Resonance Fingerprinting (MRF) has emerged as a promising quantitative imaging technique within the field of Magnetic Resonance Imaging (MRI), offers comprehensive insights into tissue properties by simultaneously acquiring multiple tissue parameter maps in a single acquisition. Sequence optimization is crucial for improving the accuracy and efficiency of MRF. In this work, a novel framework for MRF sequence optimization is proposed based on the Ziv-Zakai bound (ZZB). Unlike the Cramér-Rao bound (CRB), which aims to enhance the quality of a single fingerprint signal with deterministic parameters, ZZB provides insights into evaluating the minimum mismatch probability for pairs of fingerprint signals within the specified parameter range in MRF. Specifically, the explicit ZZB is derived to establish a lower bound for the discrimination error in the fingerprint signal matching process within MRF. This bound illuminates the intrinsic limitations of MRF sequences, thereby fostering a deeper understanding of existing sequence performance. Subsequently, an optimal experiment design problem based on ZZB was formulated to ascertain the optimal scheme of acquisition parameters, maximizing discrimination power of MRF between different tissue types. Preliminary numerical experiments show that the optimized ZZB scheme outperforms both the conventional and CRB schemes in terms of the reconstruction accuracy of multiple parameter maps.
\end{abstract}


\section{Introduction}
Magnetic resonance fingerprinting (MRF) \cite{maMagneticResonanceFingerprinting2013} is a promising accelerated quantitative magnetic resonance imaging (qMRI) technology. In contrast to conventional MRI, which relies on steady-state signals, MRF leverages transient signals generated by varying acquisition parameters, such as flip angles and repetition times. These time-varying parameters produce unique transient signals for different tissue types, enabling MRF to simultaneously extract multiple tissue parameter maps, such as $T_{1}$ and $T_{2}$, through a process known as dictionary matching. The schedule of acquisition parameters shapes the accuracy and efficiency of MRF. However, the high-dimensional nature, coupled with the simultaneous fine-tuning of hundreds or even thousands of parameters, renders the optimization of the MRF sequence exceedingly challenging.

In recent years, research on MRF sequence optimization can be broadly categorized into three aspects. Firstly, enhancing the distinctiveness of dictionary entries to improve the discrimination ability of MRF \cite{zouQuantitativeMRRelaxation2021}. Cohen et al. \cite{cohenAlgorithmComparisonSchedule2017} employed several optimization algorithms to minimize the discrepancy between the inner products of dictionary entries and the identity matrix. Sommer et al \cite{sommerPredictingEncodingCapability2017} adopted a metric that integrates the local similarity of dictionary entries and the impact of matching errors to indicate the sequence's encoding capability. Secondly, various approaches, including the application of quality factors \cite{karaParameterMapError2019}, perturbation theory \cite{stolkUnderstandingCombinedEffect2019}, and explicit first-principles simulation \cite{jordanAutomatedDesignPulse2021}, have been utilized in different studies to predict errors in MRF parameter maps and assess sequence performance. Lastly, estimation theory has been introduced to evaluate the potential of MRF sequences in identifying underlying tissue parameters. Zhao et al. \cite{zhaoOptimalExperimentDesign2019} derived the Cramér-Rao bound (CRB) to estimate the mean square error (MSE) of an unbiased parameter estimator, offering a way to evaluate the performance of optimal MRF sequence design under a high signal-to-noise ratio (SNR) conditions.

However, minimizing the CRB over a deterministic single voxel merely confines the observed signal to tightly cluster around the ideal signal, rather than directly improving the discrimination ability of MRF for different underlying tissue parameters. Therefore, CRB may not be an ideal indicator of MRF accuracy in the context of sequence optimization. The dictionary matching procedure in MRF involves matching each fingerprint signal with entries in the dictionary, which resembles the operation of a matched filter. As a result, the Ziv-Zakai bound (ZZB) \cite{chazanImprovedLowerBounds1975}, which is derived from the matched filter principles, provides a more appropriate lower bound for evaluating MRF performance. Furthermore, ZZB can leverage the \emph{a priori} information (e.g., the range or distribution of underlying tissue parameters).

In this paper, a novel framework for optimizing MRF sequence is presented. The performance of the optimized acquisition scheme is evaluated using ZZB, which incorporates knowledge of the \emph{a priori} distribution of underlying tissue parameters. Given the fundamental ZZB expression proposed in \cite{bellExtendedZivZakaiLower1997}, the explicit ZZB for the estimation of tissue parameters in MRF is derived. The optimal scheme is identified by the schedule of acquisition parameters that minimizes ZZB, i.e., maximizing discrimination power between different tissue types. Furthermore, preliminary numerical experiments validate the effectiveness of the proposed framework.

\section{Ziv-Zakai bound for MRF sequence optimization}

\subsection{Signal Model}

In this section, an MRF signal model is constructed with reference to \cite{zhaoOptimalExperimentDesign2019}. Correspondingly, each tissue voxel is assigned a univariate tissue parameter, $\theta$, with variations in $T_{1}$ while $T_{2}$ remains fixed, and vice versa. Assuming the acquisition sequence comprises $N$ time points, during which time-varying data acquisition parameters, including flip angles $\alpha$ and repetition times $TR$, are applied. Given an acquisition sequence, the transverse magnetization $m(t;\theta)$ associated with the underlying tissue parameter $\theta$ can be obtained by solving the Bloch equation, under the assumption of fully-sampled Cartesian k-space data. For simplicity, a simplified data model is established based on a single voxel, equivalent to assuming fully-sampled multi-voxel data. The observed signal from arbitrary discrete time frames of the magnetization evolution signal $s(t)$ can be modeled as
\begin{align}
   s_{n} = m[n;\theta] + z[n],\ \forall n=1,2,\dots,N,
\end{align}
where, $s_{n}$ represents the observed magnetization signal at the $n$th time frame. $m[n;\theta]$ denotes  the $n$th time frame of $m(t;\theta)$. $z[n]$ is the independent, identically distributed zero-mean additive white Gaussian noise.

By stacking the measurements as $\boldsymbol{s} =[s_{1},s_{2},...,s_{N} ]^{T}$, it can be written in vector form as 
\begin{align}
  \boldsymbol{s} = \boldsymbol{m}(\theta) + \boldsymbol{z},
\end{align}
where $\boldsymbol{m}(\theta) \in \mathbb{C} ^{N}$ denotes discrete-time representations of the transverse magnetization signal associated with the underlying tissue parameter $\theta$. And $\boldsymbol{z} \in \mathbb{C} ^{N}$ follows zero-mean Gaussian distribution $\mathcal{N} (\boldsymbol{0}, \boldsymbol{C}_{\boldsymbol{z}\boldsymbol{z}})$ and is independent of $\theta$.

For a given tissue parameter $\theta$, the measurement vector $\boldsymbol{s}$ follows a Gaussian distribution $\mathcal{N} (\bar{\boldsymbol{s}}_{\theta} , \boldsymbol{C}_{\boldsymbol{s}\boldsymbol{s}})$ with the mean vector $\bar{\boldsymbol{s}}_{\theta} = \boldsymbol{m}(\theta)$, and the covariance matrix $\boldsymbol{C}_{\boldsymbol{s}\boldsymbol{s}} = \boldsymbol{C}_{\boldsymbol{z}\boldsymbol{z}}$. Correspondingly, the conditional probability density function (PDF) of the measurement vector $\boldsymbol{s}$ given tissue parameter $\theta$ can be expressed as
\begin{align}
   f(\boldsymbol{s}\vert \theta) = 
 \frac{1}{(2\pi) ^{\frac{N}{2}}\left\lvert \boldsymbol{C}_{\boldsymbol{s}\boldsymbol{s}}\right\rvert^{\frac{1}{2}}}
 e^{-\frac{1}{2}(\boldsymbol{s}-\bar{\boldsymbol{s}}_{\theta})^{T} \boldsymbol{C}_{\boldsymbol{s}\boldsymbol{s}}^{-1}(\boldsymbol{s}-\bar{\boldsymbol{s}}_{\theta})}
 \label{eq:pdf}.
\end{align}

\subsection{ZZB Derivation}
For any estimator $\hat{\theta}(\boldsymbol{s})$, which estimates the continuous scalar random variable $\theta$ based upon noisy observations $ \boldsymbol{s}$, the estimation error is $ \epsilon_{\theta} = \hat{\theta}(\boldsymbol{s})-\theta$. The MSE is defined as
\begin{align}
  \mathrm{MSE}=\mathbb{E}\{|\hat{\theta}(\boldsymbol{s})-\theta |^{2}\} = \overline{\epsilon^{2}_{\theta}}.
\end{align}
Mathematically, ZZB relates the MSE to the probability of error of the binary hypothesis testing problem
\begin{equation}
\begin{aligned}
  & \mathcal{H} _{0}:\theta=\varphi, \;\;\qquad \mathrm{Pr}(\mathcal{H} _{0}) =\frac{f_{\theta}(\varphi)}
  {f_{\theta}(\varphi)+f_{\theta}(\varphi+\xi)}\\
  &\mathcal{H} _{1}:\theta=\varphi+\xi, \quad \mathrm{Pr}(\mathcal{H} _{1})=1-\mathrm{Pr}(\mathcal{H} _{0}).
\end{aligned}
\end{equation}
And the scalar ZZB lower bounds the MSE as \cite{chazanImprovedLowerBounds1975}
\begin{align}
  &\overline{\epsilon^{2}_{\theta}}\!\geq\!\!\frac{1}{2}\!\int_{0}^{\delta }\!\!\!\xi\!\!\int_{\vartheta\! _{\mathrm{min}}}^{\vartheta\! _{\mathrm{max}}\!-\!\xi}\!\![  f_{\theta}(\varphi)\!\!+\!\!f_{\theta}(\varphi\!+\!\xi)]
\mathrm{P\!_{min}}(\varphi, \varphi\!+\!\xi) d\varphi d\xi,\!
\end{align}
where $\theta \in [\vartheta _{\mathrm{min}}, \vartheta _{\mathrm{max}}]$, and $\delta = \vartheta _{\mathrm{max}} - \vartheta _{\mathrm{min}}$ denotes the range of univariate parameter $\theta$. $f_{\theta}(\varphi)\triangleq f(\theta)\vert_{\theta=\varphi}$ with $f(\theta)$ represents a priori PDF of  $\theta$. And $\mathrm{P_{min}}(\varphi, \varphi +\xi)$ denotes the minimum probability of error via the maximum a posteriori probability (MAP) criterion.
\begin{equation}
\begin{aligned}
  \mathrm{P}&\mathrm{_{min}}(\varphi, \varphi+\xi)\\  
  & \triangleq\mathrm{Pr}(\mathcal{H} _{0})\mathrm{Pr}(\mathcal{L} _{\varphi}\leq \gamma)
   +\mathrm{Pr}(\mathcal{H} _{1})\mathrm{Pr}(\mathcal{L} _{\varphi+\xi}> \gamma),
\end{aligned}
\end{equation}
where $\mathrm{Pr}(\,\cdot\,)$ is the probability, 
\begin{align} 
  \mathcal{L} _{\theta} = \mathrm{ln}\frac{f(\boldsymbol{s}\vert \varphi)}{f(\boldsymbol{s}\vert \varphi+\xi)}
\end{align} denotes the log-likelihood ratio (LLR) test and 
\begin{align} 
  \gamma = \mathrm{ln}\frac{\mathrm{Pr}(\mathcal{H} _{1})}{\mathrm{Pr}(\mathcal{H} _{0})}.
\end{align}

Using \eqref{eq:pdf}, the expression of LLR test can be obtained as
\begin{equation}
\begin{aligned}
\mathcal{L} _{\theta} = 
\frac{1}{2}\big[(\boldsymbol{s}-\bar{\boldsymbol{s}}_{\varphi+\xi})^{T}
\boldsymbol{C}_{\boldsymbol{s}\boldsymbol{s}}^{-1}
(\boldsymbol{s}-\bar{\boldsymbol{s}}_{\varphi+\xi}) \\
-(\boldsymbol{s}-\bar{\boldsymbol{s}}_{\varphi})^{T}
\boldsymbol{C}_{\boldsymbol{s}\boldsymbol{s}}^{-1}
(\boldsymbol{s}-\bar{\boldsymbol{s}}_{\varphi})\big],
\end{aligned}
\end{equation}
where $\bar{\boldsymbol{s}}_{\varphi}\triangleq\bar{\boldsymbol{s}}_{\theta=\varphi} =\boldsymbol{m}(\varphi)$, and thus $ \mathcal{L} _{\theta}$ is a function of underlying tissue parameter $\theta$ via $ \boldsymbol{s}$. To express $ \mathrm{Pr}(\mathcal{L} _{\varphi}\leq \gamma)$ and $ \mathrm{Pr}(\mathcal{L} _{\varphi+\xi} > \gamma)$ in explicit form, it is essential to locate the distributions of $\mathcal{L} _{\varphi}$ and $ \mathcal{L} _{\varphi+\xi}$. Given $ \theta = \varphi$, this yields
\begin{equation}
\begin{aligned}
  \mathcal{L} _{\varphi}=\frac{1}{2}\big[&(\bar{\boldsymbol{s}}_{\varphi}-\bar{\boldsymbol{s}}_{\varphi+\xi})^{T}
\boldsymbol{C}_{\boldsymbol{s}\boldsymbol{s}}^{-1}
(\bar{\boldsymbol{s}}_{\varphi}-\bar{\boldsymbol{s}}_{\varphi+\xi})\big]\\
&+(\bar{\boldsymbol{s}}_{\varphi}-\bar{\boldsymbol{s}}_{\varphi+\xi})^{T}
\boldsymbol{C}_{\boldsymbol{s}\boldsymbol{s}}^{-1}\boldsymbol{z},
\end{aligned}
\end{equation}
where $ \boldsymbol{s}_{\varphi}\triangleq \boldsymbol{s}| _{\theta=\varphi}=\boldsymbol{m}(\varphi)+\boldsymbol{z}$. Since only the Gaussian noise $\boldsymbol{z}$ belongs to the random variable in $ \mathcal{L} _{\varphi}$,  $ \mathcal{L} _{\varphi}$ follows a Gaussian distribution $\mathcal{N} (\mu_{\mathcal{L} _{\varphi}}, \sigma _{\mathcal{L} _{\varphi}}^{2})$ with 
\begin{equation}
\begin{aligned}
  \mu_{\mathcal{L}_{\varphi}}=\frac{1}{2}[\boldsymbol{m}(\varphi)-\boldsymbol{m}(\varphi+\xi)]^{T}
  \boldsymbol{C}_{\boldsymbol{s}\boldsymbol{s}}^{-1}
  [\boldsymbol{m}(\varphi)-\boldsymbol{m}(\varphi+\xi)]
\end{aligned}
\end{equation}
and 
\begin{align}
  \sigma _{\mathcal{L} _{\varphi}}^{2}= 2\mathbb{V}\Big[(\bar{\boldsymbol{s}}_{\varphi}-\bar{\boldsymbol{s}}_{\varphi+\xi})^{T}
  \boldsymbol{C}_{\boldsymbol{s}\boldsymbol{s}}^{-1}\boldsymbol{z}\Big],
\end{align}
where $ \mathbb{V}\{ \cdot\}$ denotes variance. Given that $\boldsymbol{z} \sim \mathcal{N} (\boldsymbol{0}, \boldsymbol{C}_{\boldsymbol{z}\boldsymbol{z}})$, it follows that $  \sigma _{\mathcal{L} _{\varphi}}^{2}=2\mu_{\mathcal{L}_{\varphi}}$, as referenced in \cite{zhangZivZakaiBoundCompressive2022}. Thus, $\mathcal{L} _{\varphi}$ follows the deterministic distribution of $\mathcal{N} (\mu_{\mathcal{L}_{\varphi}}, 2\mu_{\mathcal{L}_{\varphi}})$.

Similarly, given $\theta=\varphi+\xi$, $ \mathcal{L} _{\varphi+\xi}$ follows $\mathcal{N} (-\mu_{\mathcal{L}_{\varphi}}, 2\mu_{\mathcal{L}_{\varphi}})$. Then, $\mathrm{P_{min}}(\varphi, \varphi +\xi)$ becomes
\begin{equation}
  \begin{aligned}
  \mathrm{P}&\mathrm{_{min}}(\varphi, \varphi+\xi)=\\
  &\mathrm{Pr}(\mathcal{H} _{0})
  \mathcal{Q} \Bigg(\frac{\mu_{\mathcal{L}_{\varphi}}-\gamma}
  {\sqrt{2\mu_{\mathcal{L}_{\varphi}}}}\Bigg)
  +\mathrm{Pr}(\mathcal{H} _{1})
  \mathcal{Q} \Bigg(\frac{\mu_{\mathcal{L}_{\varphi}}-\gamma}
  {\sqrt{2\mu_{\mathcal{L}_{\varphi}}}}\Bigg),
  \label{eq:Pmin}
  \end{aligned}
\end{equation}
where $\mathcal{Q}(\,\cdot\,)$ denotes the tail distribution function of the standard normal distribution.
Moreover, assuming that $\theta$ follows a uniform distribution in $[\vartheta _{\mathrm{min}}, \vartheta _{\mathrm{max}}]$, i.e., $f_{\theta}(\varphi)=f_{\theta}(\varphi+\xi)=\frac{1}{\delta}$, \eqref{eq:Pmin} becomes
\begin{align}
  \mathrm{P_{min}}(\varphi, \varphi+\xi)
  =\mathcal{Q} \Bigg(\sqrt{\frac{\mu_{\mathcal{L}_{\varphi}}}{2}}\Bigg).
\end{align}

Finally, with the assumption of white noise, i.e., $\boldsymbol{C}_{\boldsymbol{z}\boldsymbol{z}}=\sigma^{2}\boldsymbol{I}$ with noise power $\sigma^{2}$, the final bound yields
\begin{align}
  \overline{\epsilon^{2}_{\theta}}\!\geq\! \frac{1}{\delta}\!\int_{0}^{\delta }\!\xi\!
  \int_{\vartheta _{\mathrm{min}}}^{\vartheta _{\mathrm{max}}\!-\!\xi}\!
  \mathcal{Q}\Bigg(\!\sqrt{\frac{1}{4\sigma^{2}}\!
  \left\lVert\boldsymbol{m}(\varphi)\!-\!\boldsymbol{m}(\varphi\!+\!\xi)\right\rVert_{2}^{2} }\Bigg)
  \!d\varphi d\xi,
\end{align}
where $\left\lVert\,\cdot\,\right\rVert_{2}$ represents the $ \ell_{2}$ norm.

\subsection{Optimal Experiment Design}

In this paper, MRF sequence optimization is executed with the IR-FISP sequence \cite{jiangMRFingerprintingUsing2015}. Since the derived ZZB provides a lower bound on the minimum probability of error in a binary hypothesis testing problem concerning the identification of distinct tissue-specific signals, it can be used to indicate the tissue discrimination capability of the candidate sequences. Mathematically, the optimal experiment design problem is formulated as follows:
\begin{equation}
  \label{design problem}
\begin{aligned}
    \min_{\{\alpha_{n},TR_{n}\}_{n=1}^{N}} &\sum_{l = 1}^{L} W_{l} \cdot \mathrm{ZZB}(\theta^{(l)})\\[0.5mm]
    \mathrm{s.t.} \;\;\quad &TR_{n}^{\mathrm{min}} \leq TR_{n} \leq TR_{n}^{\mathrm{max}}, \quad 1\leq n\leq N,\\[0.5mm]
    &\alpha_{n}^{\mathrm{min}} \leq \alpha_{n} \leq \alpha_{n}^{\mathrm{max}}, \quad 1\leq n\leq N,\\[0.5mm]
    &| \alpha_{n+1}-\alpha_{n}| \leq \Delta \alpha_{n}^{\mathrm{max}}, \quad 1\leq n\leq N-1,
\end{aligned}
\end{equation}
where the collection of representative scalar tissue parameters $\{\theta^{(l)}\}_{l=1}^{L}$, each with respectively specified distribution range $\theta^{(l)}\in [\vartheta ^{(l)}_{\mathrm{min}},\vartheta ^{(l)}_{\mathrm{max}}]$, is incorporated into the experimental design. $N$ is the predefined total number of time points, and $\{W_{l}\}_{l=1}^{L}$ denotes manually selected weighting parameters that balance the importance of different tissue parameters. $ZZB(\,\cdot\,)$ denotes the ZZB value obtained for a given scalar tissue parameter $\theta^{(l)}$. Furthermore, constraints are introduced to bound the acquisition parameters at each time point, i.e., $TR_{n} \in [TR_{n}^{\mathrm{min}},TR_{n}^{\mathrm{max}}]$ and $\alpha_{n} \in [\alpha_{n}^{\mathrm{min}},\alpha_{n}^{\mathrm{max}}] $, and restrict the maximum flip angle variations between consecutive time points within $\Delta \alpha_{n}^{\mathrm{max}}$. Here, the optimization problem \eqref{design problem} was addressed using the sequential least-square quadratic programming (SLSQP) algorithm.

\section{Implementation and results}
\begin{figure}[htb]
  \centering
  \begin{minipage}[b]{.1\linewidth}
    \centering
    \centerline{\includegraphics[width=1.58cm]{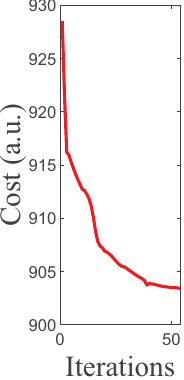}}
    \centerline{(a)}
    \medskip
  \end{minipage}
  \begin{minipage}[b]{.52\linewidth}
  \centering
  \centerline{\includegraphics[width=3.0cm]{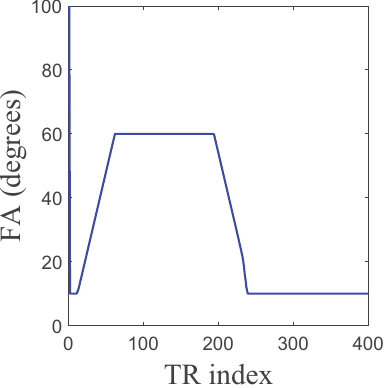}}
  \centerline{(b)}
  \medskip
  \end{minipage}
  \begin{minipage}[b]{.2\linewidth}
  \centering
  \centerline{\includegraphics[width=3.0cm]{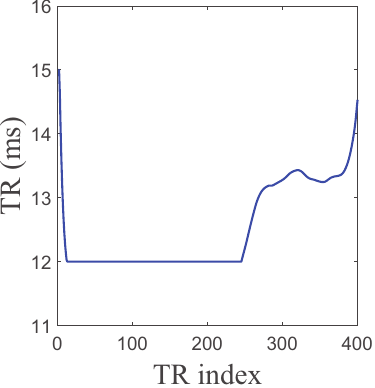}}
  \centerline{(c)}
  \medskip
  \end{minipage}
  \hfill
  \caption{Optimization cost curve and the optimized FA/TR scheme. (a): Optimization cost as a function of the number of iteration. (b)-(c): FA/TR train of the optimized scheme.}
  \label{fig:op-seq}
  \end{figure}

In this section, the detailed implementation of the proposed sequence optimization framework is presented in accordance with \eqref{design problem}. Specifically, Table \ref{tab:tissue pairs} lists ten pairs of selected representative tissue parameters, referencing the fundamental components of the brain outlined in \cite{nagtegaalEstimationMultipleComponents2023}, along with their respective weights. Moreover, let $N=400$, $[\alpha_{1}^{\mathrm{min}}, \alpha_{1}^{\mathrm{max}}]=[10^\circ,180^\circ]$ with $\Delta \alpha _{1}^{\mathrm{max}} = +\infty$ for $\alpha_{1}$, $[\alpha_{n}^{\mathrm{min}}, \alpha_{n}^{\mathrm{max}}]=[10^\circ,60^\circ]$ with $\Delta \alpha _{n}^{\mathrm{max}} = 1^\circ$ for other flip angles, and $[TR_{n}^{\mathrm{min}}, TR_{n}^{\mathrm{max}}]=[12\mathrm{ms},15\mathrm{ms}]$. In addition, the SLSQP algorithm is employed, initiating from the truncated conventional scheme \cite{jiangMRFingerprintingUsing2015}, to solve the optimization problem associated with \eqref{design problem}. The algorithm is terminated till the relative change in the cost for successive iterations is less than a predefined tolerance value. The runtime of the algorithm took about 120 min on a Linux workstation with an Intel Xeon 3.80 GHz CPU and an Nvidia Quadro GV100 GPU. Finally, the optimization cost curve is shown in Fig.\ref{fig:op-seq}(a). Fig.\ref{fig:op-seq}(b)-(c) exhibits the optimized acquisition parameters, referred as the ZZB scheme for convenience.

\begin{figure*}[htb]
  \centering
  \includegraphics[width=0.63\linewidth]{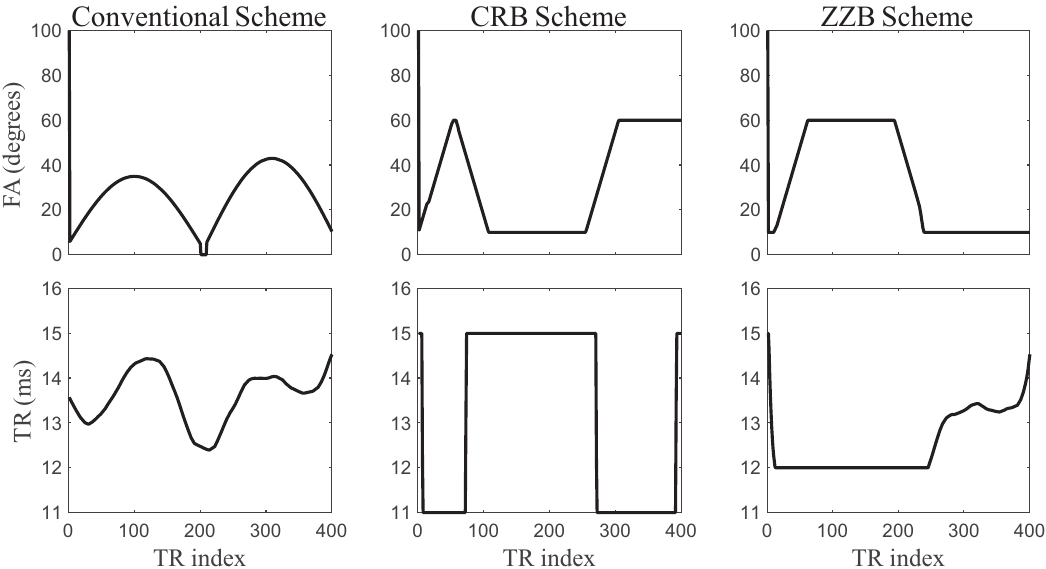}
  \caption{Acquisition parameter schemes. The columns, from left to right, show the conventional scheme, CRB scheme, and ZZB scheme.}
  \label{fig:compare_schemes}
  \end{figure*}

\begin{figure*}[htb]
  \centering
  \includegraphics[width=0.8\linewidth]{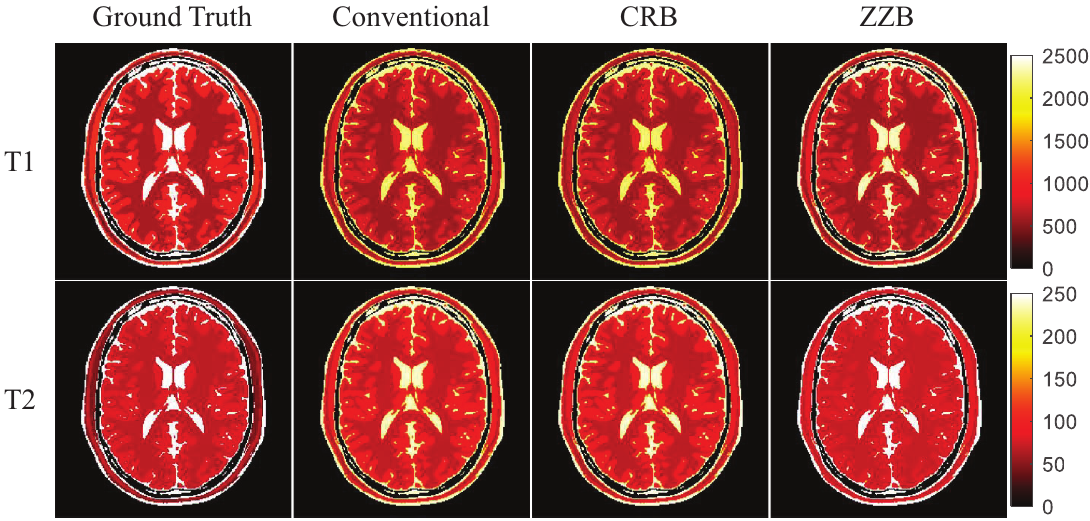}
  \caption{Reconstructed parameter maps. From the left to the right columns are the ground truth maps, reconstructed maps using different acquisition schemes.}
  \label{fig:recon para}
  \end{figure*}

\begin{figure*}[htb]
  \centering
  \hspace{2.5cm} 
  \includegraphics[width=0.6\linewidth]{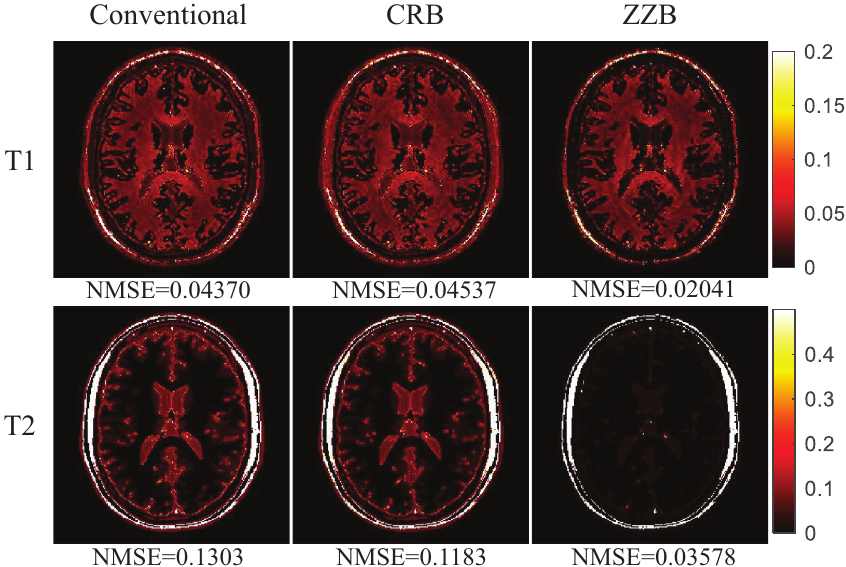}
  \caption{Error maps between reconstructed parameter maps using different acquisition schemes and the ground truth maps.}
  \label{fig:para error}
\end{figure*}

To assess the effectiveness of the proposed framework, simulation studies were conducted using a numerical brain phantom with ground $T_{1}$ and $T_{2}$ maps form the Brainweb database \cite{collinsDesignConstructionRealistic1998}, as depicted in the first column of Fig.\ref{fig:recon para}. In the Bloch simulation, MRF acquisition was simulated using the conventional, CRB \cite{zhaoOptimalExperimentDesign2019}, and ZZB schemes individually, as illustrated in Fig.\ref{fig:compare_schemes}. Each simulation maintained a consistent acquisition length of $N=400$, a fixed echo time $TE=3\mathrm{ms}$, and utilized the same highly-undersampled spiral trajectory as \cite{jiangMRFingerprintingUsing2015}. The 2nd through 4th columns of Fig.\ref{fig:recon para} depict the reconstructed parameter maps of $T_{1}$ and $T_{2}$ via low-rank approach \cite{mazorLowrankMagneticResonance2018,zhaoImprovedMagneticResonance2018,huHighQualityMRFingerprinting2022} with different acquisition schemes, respectively. Additionally, the normalized mean square error (NMSE) of the reconstructed parameter maps was evaluated. To emphasize differences in the quality of the reconstructed parameter maps, Fig.\ref{fig:para error} illustrates the corresponding error map for clearer contrast. The ZZB scheme demonstrates a significant enhancement in the $T_{2}$ reconstruction accuracy, accompanied by a moderate improvement in $T_{1}$ reconstruction accuracy compared to both the conventional and CRB schemes.

\begin{table*}[htbp]  
  \centering  
  \caption{Ten pairs of selected tissue parameters and corresponding weights.}
  \resizebox{\linewidth}{!}{
  \begin{tabular}{ccccccccccccc}
    \bottomrule
    No. & 1 & 2 & 3 & 4 & 5 & 6 & 7 & 8 & 9 & 10\\
    \hline
    T1(ms) & [100,500] & [500,2000] & [700,1100] & [1200,1600] & [2000,4000]& 150 & 1800 & 900 & 1500 & 3500\\
    T2(ms) & 20 & 800 & 60 & 90 & 400 & [10,20] & [200,1200] & [50,80] & [60,120] & [300,2000]\\
    Weights & 0.0050 & 0.0005 & 0.0100 & 0.0100 & 0.0002 & 5.0000 & 0.0005 & 2.0000 & 0.5000 & 0.0002\\
    \bottomrule
  \end{tabular}}
  \label{tab:tissue pairs}
\end{table*} 

\section{Conclusion}

In this paper, a novel framework for optimizing MRF sequence is proposed based on ZZB. ZZB provides a fresh perspective to quantify the minimum achievable discrimination error for candidate sequences in MRF. The optimal experiment design problem was formulated to maximize the discrimination power, which directly correlates with the accuracy of MRF, in the resulting optimized acquisition scheme. Preliminary numerical results demonstrate that the optimized ZZB scheme significantly enhances the reconstruction accuracy of the $T_{2}$ map and moderately improves the reconstruction accuracy of the $T_{1}$ map. Additionally, the ZZB scheme outperforms both the conventional and CRB schemes based on the current simulations. In the future, incorporating the bias-variance property of the reconstructed parameter maps will contribute to further refining the theoretical analysis of the proposed framework.

\footnotesize \bibliographystyle{IEEEbib}
\bibliography{myref}

\clearpage

\end{document}